\documentclass[twocolumn,letter]{jpsj3} %% two-column layout
\usepackage{graphicx}% Include figure files
\usepackage{dcolumn}% Align table columns on decimal point
\usepackage{bm}% bold math
\usepackage{color}

\title{Pressure-Temperature-Magnetic Field Phase Diagram \\
of Ferromagnetic Kondo Lattice CeRuPO}

\author{Hisashi Kotegawa$^{1}$, Toshihiro Toyama$^1$, Shunsaku Kitagawa$^1$, Hideki Tou$^1$ \\
Ryota Yamauchi$^1$, Eiichi Matsuoka$^1$, and Hitoshi Sugawara$^1$}

\inst{$^1$Department of Physics, Kobe University, Kobe 657-8501, Japan

}

\abst{
We report the temperature-pressure-magnetic field phase diagram made from electrical resistivity measurements for the ferromagnetic (FM) Kondo lattice CeRuPO.
The ground state at zero field changes from the FM state to another state, which is suggested to be an antiferromagnetic (AFM) state, above $\sim0.7$ GPa, and the magnetically ordered state is completely suppressed at $\sim2.8$ GPa.
In addition to the collapse of the AFM state under pressure and a magnetic field, a metamagnetic (MM) transition from a paramagnetic state to a polarized paramagnetic state appears.
CeRuPO will give us a rich playground for understanding the mechanism of the MM transition under comparable FM and AFM correlations in the Kondo lattice.
}

\kword{CeRuPO, ferromagnetic Kondo lattice, metamagnetism, pressure}

\begin{document}
\maketitle

The metamagnetic (MM) transition in correlated electron systems has been an interesting subject related to magnetic instability and Fermi surface instability.
An excellent  example is the heavy-fermion system CeRu$_2$Si$_2$ and doped systems.\cite{Haen,Flouquet,Knafo,Aoki_CeRu2Si2,Shimizu,Haen2,Matsumoto,Sato,HAoki,Daou,Aoki_Rh}
In the antiferromagnetic (AFM) system with Rh, La, and Ge doping, a magnetic field suppresses the ordered phase in the critical field, inducing the MM transition.\cite{Flouquet,Knafo,Aoki_CeRu2Si2,Shimizu,Haen2,Matsumoto}
Even in the absence of the AFM phase, if the system is located close to the AFM instability, the proximity of the AFM critical field yields the MM transition.
In addition to the field evolution of the AFM correlation, the presence of the ferromagnetic (FM) correlation also plays a vital role in the MM transition in CeRu$_2$Si$_2$,\cite{Sato} and the FM state is realized on the Ge-rich side in CeRu$_2$(Si,Ge)$_2$.\cite{Haen2,Matsumoto}
Another key factor for the MM transition is a change in the Fermi surface related to the breakdown of the Kondo effect.
If the Kondo temperature $T_K$ is low, a magnetic field corresponding to $T_K$ can quench the Kondo effect, inducing the MM transition.
The interpretation of the change in the Fermi surface is still a subject of debate.\cite{HAoki,Daou}
In Rh-doped CeRu$_2$Si$_2$, a clear separation of two MM transitions demonstrates the presence of two mechanisms for the MM transition.\cite{Aoki_Rh}
On the other hand, in the system with the FM instability, the MM transition from the paramagnetic (PM) state to the FM state is realized because of strong FM correlations.
If the system has a tricritical point (TCP) where the second-order phase transition at Curie temperature $T_C$ changes into the first-order phase transition, a wing structure of the first-order MM transition appears in the pressure($P$)-temperature($T$)-magnetic field($H$) phase diagram.\cite{Belitz,Yamada}
%The quantum critical endpoint (QCEP), where the first order MM transition termi%nates at  0 K, is a fascinating subject as a new type of quantum critical point% (QCP).\cite{Millis,Imada,Yamaji}
Such a phase diagram can be explained in the framework of itinerant FM systems and has been confirmed in $5f$ systems such as UGe$_2$\cite{Valentin,Kotegawa} and UCoAl\cite{Aoki} and $3d$ systems such as ZrZn$_2$.\cite{Kabeya}

CeRuPO is a rare example of the FM Kondo lattice among $4f$ electron systems.\cite{Krellner,Krellner2}
Its FM ordered moments are aligned along the $c$-axis in the tetragonal structure below $T_C=14.5$ K, although a larger magnetization is induced along the $ab$ plane in high-temperature or high-field regions owing to the effect of a crystal electric field (CEF).\cite{Krellner2}
The Kondo effect has been confirmed in resistivity, specific heat, and thermoelectric power measurements.\cite{Krellner}
Therefore, the FM ordered state is expected to be suppressed by some tuning parameter as in the case of AFM Kondo lattices.
In fact, Kitagawa {\it et al.} have found that the substitution of Ru by Fe suppresses $T_C$ towards 0 K,\cite{Kitagawa2} although they have interpreted that the suppression of the FM state is not induced by the enhancement of the Kondo effect but by the change in the dimensionality of magnetic correlations.\cite{Kitagawa3}
In this substitution system, it has been reported that an FM quantum critical point (QCP) appears without a distinct signature of the TCP.
The MM crossover in the PM regime is found under the magnetic field along the $ab$ plane.\cite{Kitagawa2,Kitagawa1}
On the other hand, the use of a pressure application is a promising way to suppress the ordered state without giving much inhomogeneity.
Macovei {\it et al.} have performed a resistivity measurement under pressure of up to $\sim2.3$ GPa, confirming the suppression of the ordered state.\cite{Macovei}
They estimated the critical pressure of the FM instability to be $\sim3.2$ GPa from the extrapolation of the phase diagram, although they did not attain it.
The aims of this study are to determinate the phase diagram in a higher pressure region and to search for the MM transition in CeRuPO. 
This will give us a good opportunity to obtain the $P-T-H$ phase diagram for the 4$f$ FM Kondo lattice in contrast to more itinerant 5$f$ and 3$d$ systems, as well as to understand the magnetic character of CeRuPO.

Single crystals of CeRuPO and CeFePO were prepared by the Sn-flux method described in Ref.~19.
Current was made to flow along the $ab$ plane to measure the electrical resistivity $\rho$, and the measurements under magnetic field were performed in the longitudinal configuration for $H \parallel ab$.
An indenter-type pressure cell and Daphne7474 as the pressure-transmitting medium were utilized for pressure experiments.\cite{Indenter,Murata}
The pressure at a low temperature was estimated using a Pb manometer.

\begin{figure}[htb]
\centering
\includegraphics[width=0.8\linewidth]{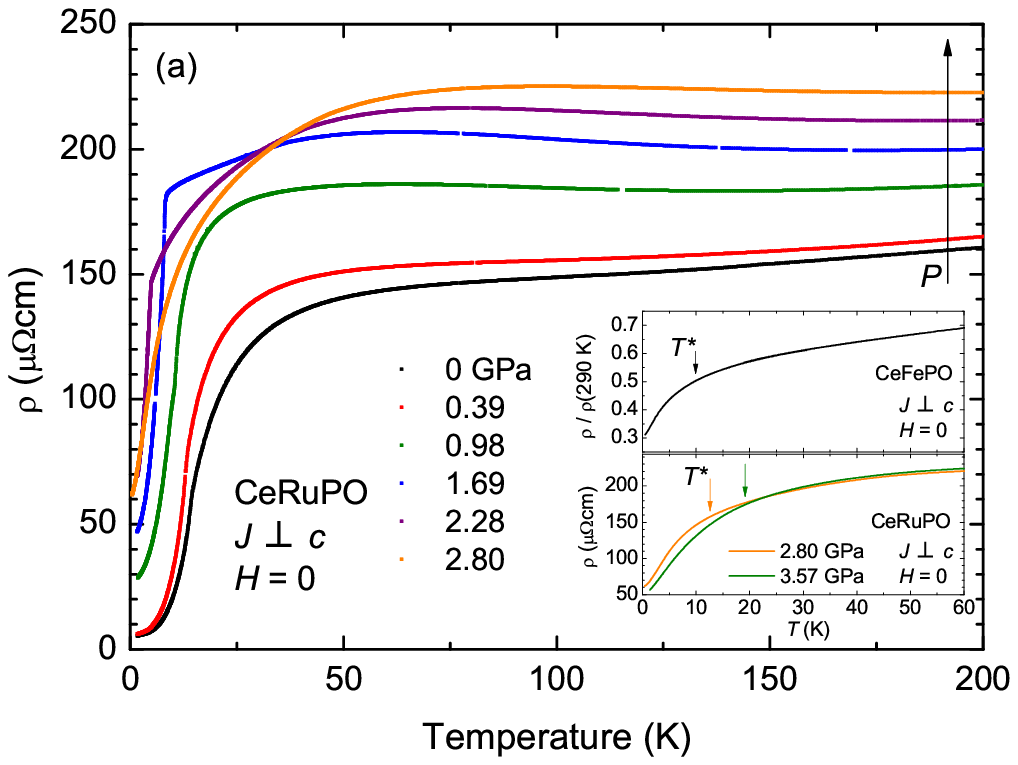}
\includegraphics[width=0.8\linewidth]{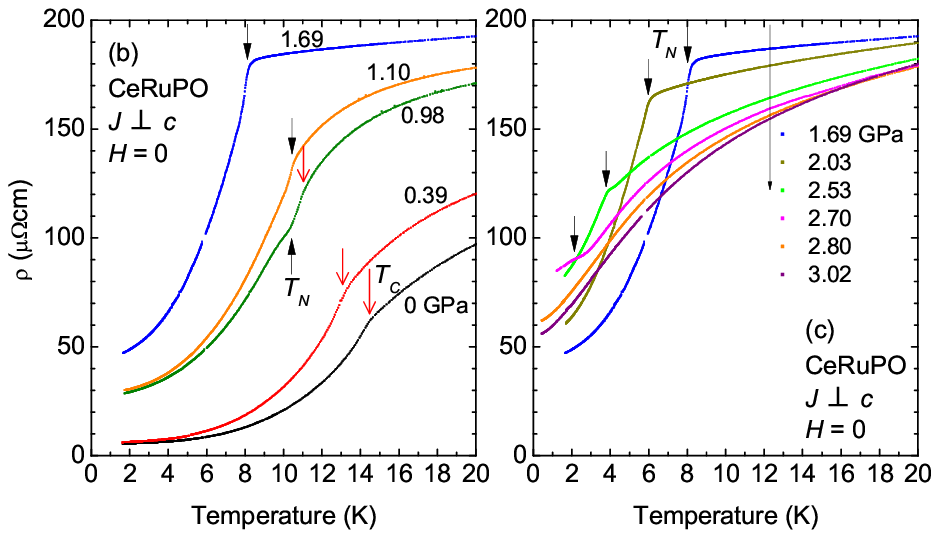}
\caption[]{(color online) Temperature dependences of $\rho$ for CeRuPO at several pressures and zero field. The inset shows a comparison between CeRuPO under pressure and CeFePO, where the coherent temperature $T^*$ indicated by arrows was estimated from a kink in $d^2\rho(T) / dT^2$. $T_C$ decreases with increasing pressure, and a successive transition appears at 0.98 GPa. The second transition is also suppressed under pressure and disappears above 2.80 GPa.
}
\end{figure}

Figures 1(a)-1(c) show the temperature dependences of $\rho$ measured at several pressures.
The shoulder in $\rho(T)$ seen at approximately 30 K once decreases with increasing pressure and then increases above 1 GPa, as in a previous report.\cite{Macovei}
$T_{\rm C}$ decreases gradually with increasing pressure, and a successive anomaly appears at 0.98 GPa as indicated by two types of arrow in Fig.~1(b).
The clear hysteresis convinces us that the second anomaly is a first-order phase transition (not shown).
The successive transition is seen in a narrow pressure region and only one transition is observed above 1.10 GPa.
As shown in Figs.~1(b) and 1(c), the transition temperature above 1.10 GPa also decreases with increasing pressure, and the anomaly disappears above 2.80 GPa.
The inset in Fig.~1(a) shows a comparison between CeRuPO under pressure and CeFePO.
CeRuPO at 2.80 GPa and CeFePO are both located close to the magnetic instability.\cite{Kitagawa2,Lausberg}
In CeRuPO at 2.80 GPa, $\rho(T)$ starts to decrease below $\sim100$ K and shows a sharp decrease again below $\sim10-15$ K.
The anomaly at high temperature is considered to originate from the combination of the Kondo effect and the CEF splitting,\cite{Macovei} and the shoulder at low temperature is attributed to the development of a coherent Kondo state.
The characteristic coherent temperature $T^*$ was estimated from a kink in $d^2\rho(T) / dT^2$ and is indicated by arrows in the inset.
The $T^* \sim 10$ K estimated for CeFePO is somewhat higher than the characteristic temperature $T_{max}\sim5$ K determined from the nuclear spin relaxation rate.\cite{Kitagawa1}
The $T^*$ for CeRuPO at 2.80 GPa is slightly higher than the $T^*$ for CeFePO, and $T^*$ increases with increasing pressure, which is the typical behavior observed in the Kondo lattice.

\begin{figure}[htb]
\centering
\includegraphics[width=0.8\linewidth]{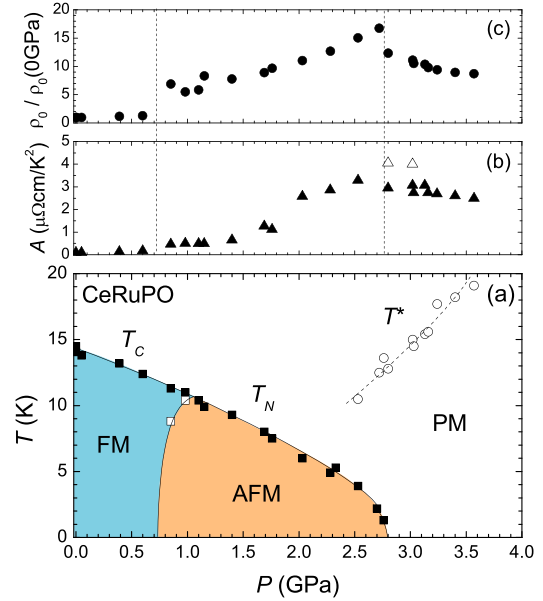}
\caption[]{(color online) (a) $P-T$ phase diagram for CeRuPO at zero field.  (b) and (c) Pressure dependences of the $A$ coefficient and residual resistivity. Both show a strong enhancement towards the critical pressure of $\sim$$2.8$ GPa. The $A$ coefficients were estimated from $\rho=\rho_0 + AT^2$ in the temperature range of $1.5-2$ K (closed triangle) and $0.4-1.4$ K (open triangle).
}
\end{figure}

\begin{figure}[htb]
\centering
\includegraphics[width=0.7\linewidth]{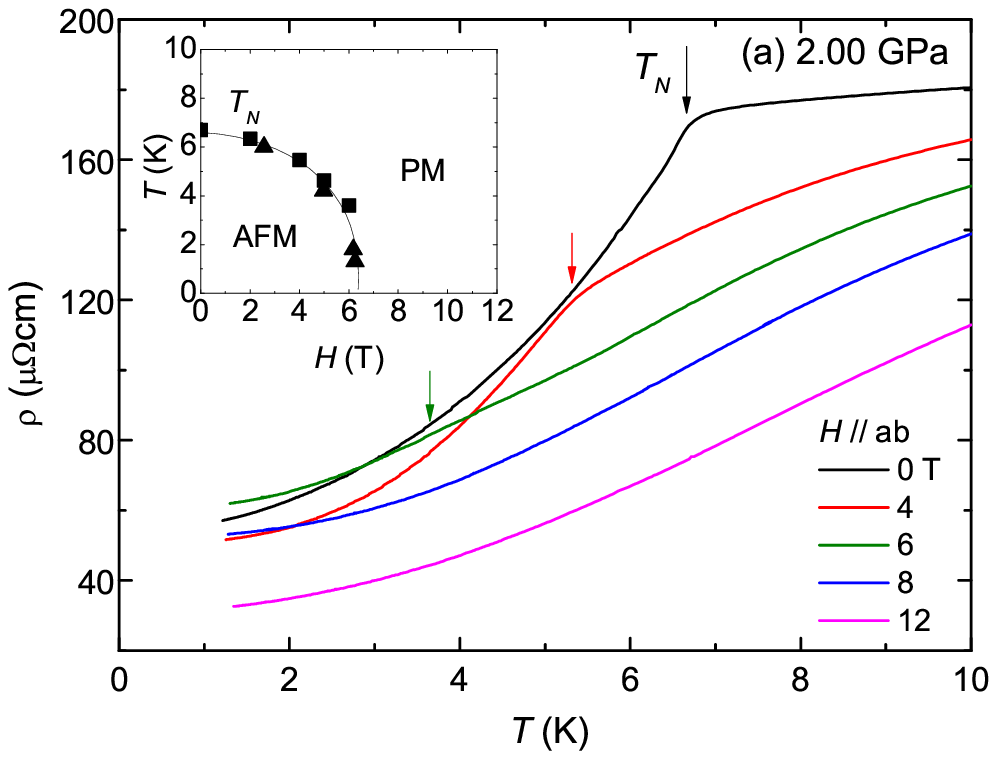} \\
\includegraphics[width=0.7\linewidth]{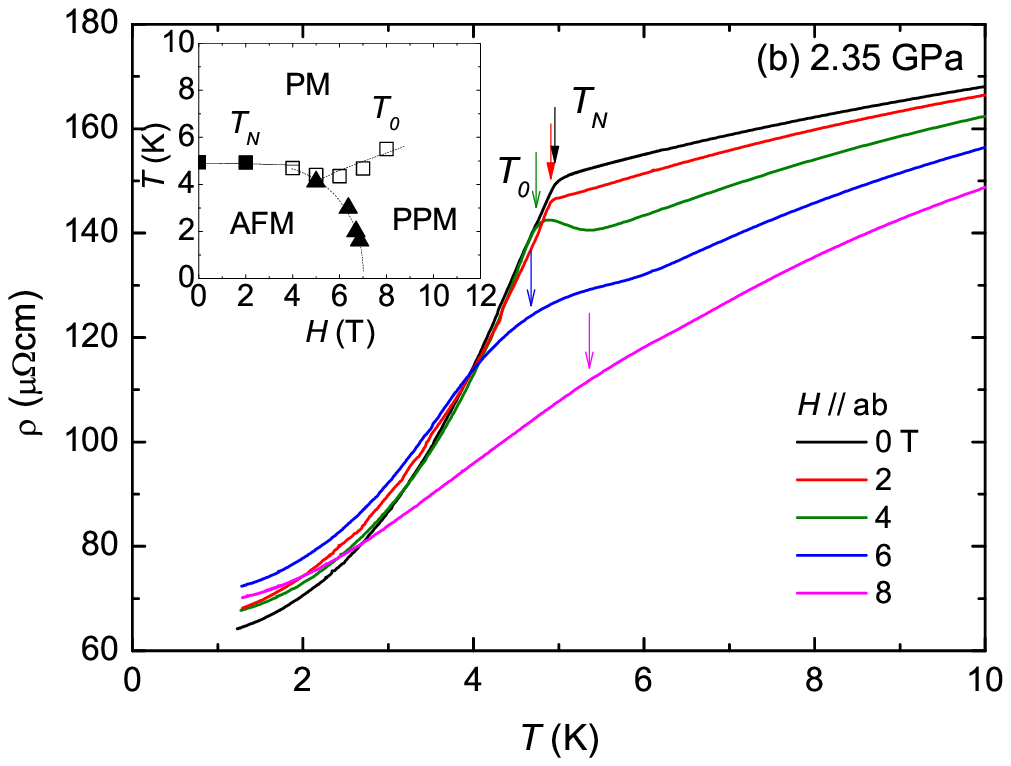} \\
\includegraphics[width=0.7\linewidth]{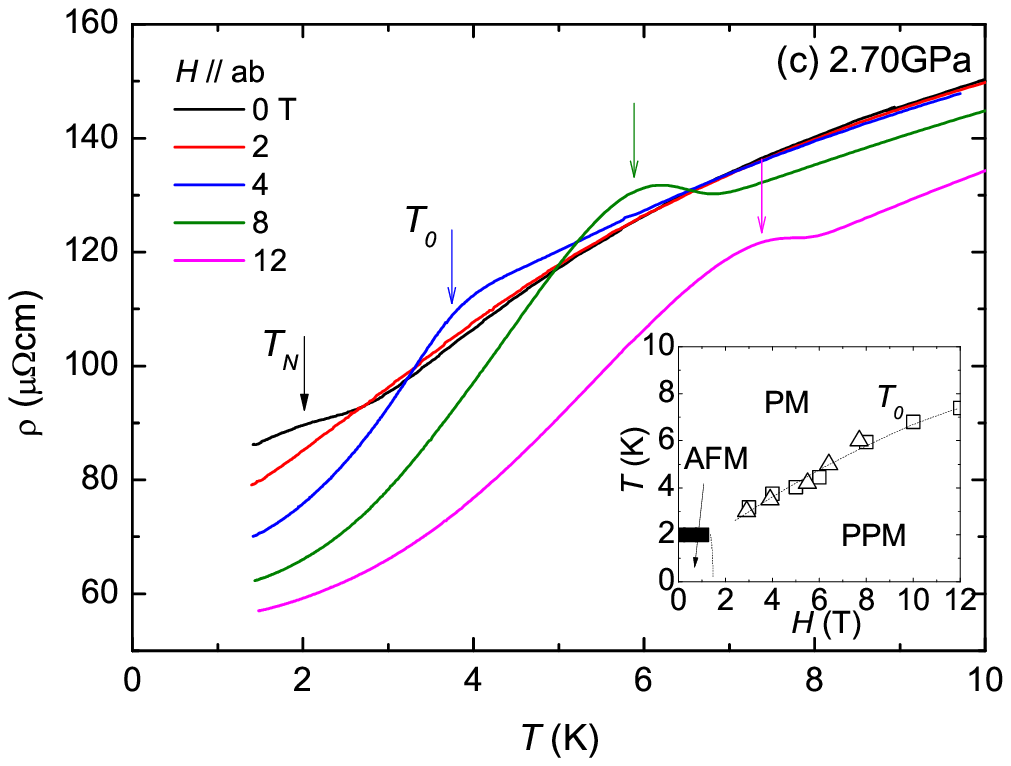} \\
\includegraphics[width=0.7\linewidth]{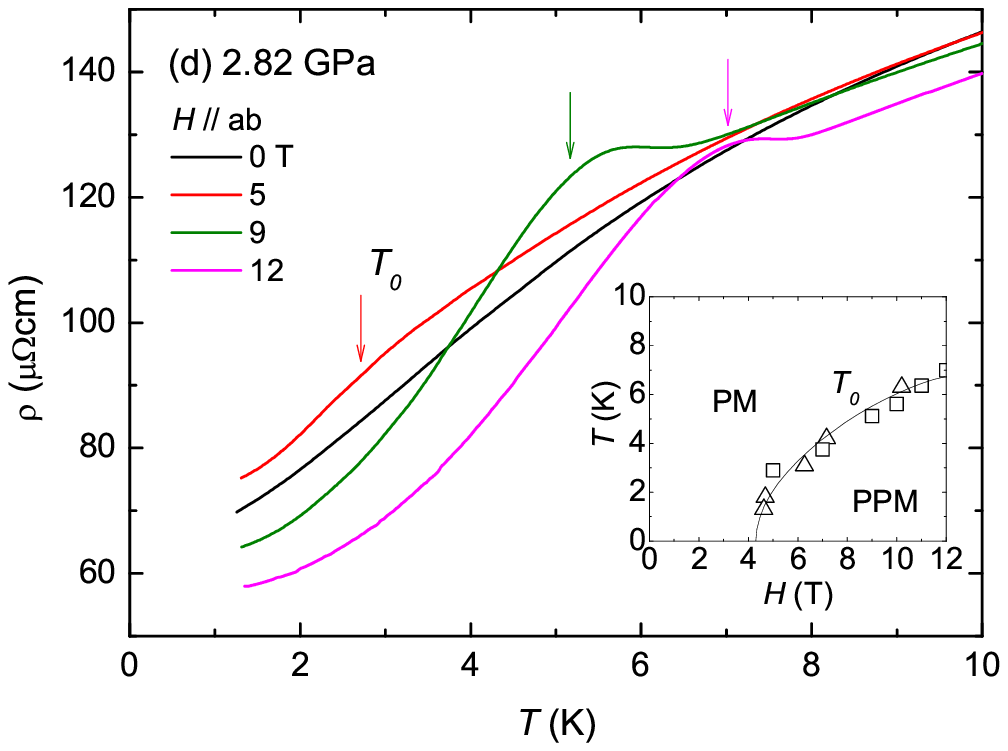}
\caption[]{(color online) (a-d) Temperature dependences of $\rho$ at different magnetic fields, which are applied along the $ab$ plane. The inset shows the $T-H$ phase diagram at each pressure.
}
\end{figure}

Figure 2(a) shows the $P-T$ phase diagram for CeRuPO at zero field.
The successive transition was observed in a narrow pressure region close to 1 GPa.
The discrete jump of the residual resistivity $\rho_0$ suggests that the ground state changes from the FM state to another at approximately 0.7 GPa.
%The ground state at ambient pressure has been reported to be the FM state where the magnetic moments are polarized along the $c$ axis.\cite{Krellner2}
The second phase under pressure is separated from the FM state by a first-order transition.
As shown in Fig.~3(a), the temperature of transition into the second phase clearly decreases under the magnetic field along the $ab$ plane.
This is in sharp contrast to the FM state at ambient pressure, where the transition temperature increases with increasing field for both $H\parallel ab$ and $H\parallel c$.\cite{Krellner2}
The suppression of the ordered phase under the magnetic field suggests that the second phase is not an FM phase but an AFM phase.
In fact, our $^{31}$P-NMR measurement under pressure shows a clear splitting of the NMR spectrum in the second phase.
This demonstrates the occurrence of the magnetic sublattice.
Therefore, we denote the second transition temperature as $T_N$ and the second phase as the AFM phase.
The detailed results of NMR measurements under pressure will appear in a separate paper.
Further pressure application completely suppresses the AFM state at approximately 2.8 GPa, accompanied by an increase in $T^*$.
The decrease in $\rho_0$ above $\sim2.8$ GPa indicates that high $\rho_0$ is inherent in the AFM phase, and that the hump anomaly at $T_N$ for 2.70 GPa is likely to originate from the increase in $\rho_0$ [see Fig.~1(b)].
We cannot determine whether the transition near 2.8 GPa is of first order or second order, but the phase diagram indicates that CeRuPO possesses the AFM instability under pressure.
This stands in contrast to the substitution system Ce(Ru,Fe)PO, where the FM state is suppressed continuously toward 0 K.\cite{Kitagawa2}
In Ce(Ru,Fe)PO, it has been reported that $q\neq0$ magnetic fluctuations develop accompanied by the suppression of the FM correlations, although the long-range ordering with a $q\neq0$ wave number is not realized.\cite{Kitagawa2}
In contrast, the AFM ordered state is realized in CeOsPO.\cite{Krellner}
These results imply that both FM and AFM correlations are present in these related systems and two correlations compete with each other in CeRuPO.
The $A$ coefficients, which are estimated from $\rho=\rho_0 + AT^2$ in the temperature range of $1.5-2$ K (closed triangle) and $0.4-1.4$ K (open triangle), are shown in Fig.~2(b).
$A$ has a strong enhancement near 2.8 GPa, indicative of the quantum criticality.
When we estimated the power $n$ in $\rho=\rho_0 + A'T^n$, it was $\sim$$1.7$ for 2.80 and 3.02 GPa by the fitting in the temperature range of $0.4-1.4$ K.

Figures 3(a)-3(d) show the temperature dependences of $\rho$ at different magnetic fields, which are applied along the $ab$ plane.
The insets show the $T$ - $H$ phase diagrams at each pressure.
At 2.00 GPa, the $T_N$ of 6.6 K decreases with the field application as mentioned above.
The kink at $T_N$ can be observed up to 6 T, but it smears out and disappears above it.
The transition temperature determined by $T$-sweep measurements is shown by squares in the inset, whereas the anomaly that appears in the $H$-sweep measurements is shown by triangles in the phase diagram.
The $H$-sweep data at $\sim1.4$ K are shown in Fig.~4.
The obtained phase diagram in the inset of Fig.~3(a) suggests that the AFM state is suppressed at $\sim6.5$ T.
At 2.35 GPa, $T_N$ first slightly decreases with increasing field, but the anomaly changes from a kink into a hump above 4 T.
The characteristic temperature of the hump is determined by the dip in $d^2\rho/dT^2$ and denoted by $T_0$.
$T_0$ increases with increasing field and the anomaly smears out at a higher magnetic field.
The increase in $T_0$ under a magnetic field is more pronounced at 2.70 GPa.
A simple consideration about free energy tells us that the increase in magnetization triggers the emergence of a high-field state, analogous to the case of conventional FM transition.
%This indicates the ground state below $T_0$ has larger magnetization along the %$ab$ plane.
%At 2.70 GPa close to the AFM instability at zero filed, the anomaly at zero fie%ld is the hump, but we classified it as the AFM transition because of the conti%nuity at zero field.
%As applying field, we cannot observe a clear anomaly at 2 T, and the anomaly de%noted by $T_0$ appears above 2 T.
This field-induced phase appears even at 2.82 GPa where the AFM state is completely suppressed, indicating that the field-induced phase is not a spin-flopped AFM phase.
The smearing of the anomaly at $T_0$ under a high field, shown in Fig.~3(b), suggests that the ground state below $T_0$ is connected with the PM state by a crossover with the same symmetry; consequently, it should be assigned to be a polarized paramagnetic (PPM) state.

\begin{figure}[htb]
\centering
\includegraphics[width=0.95\linewidth]{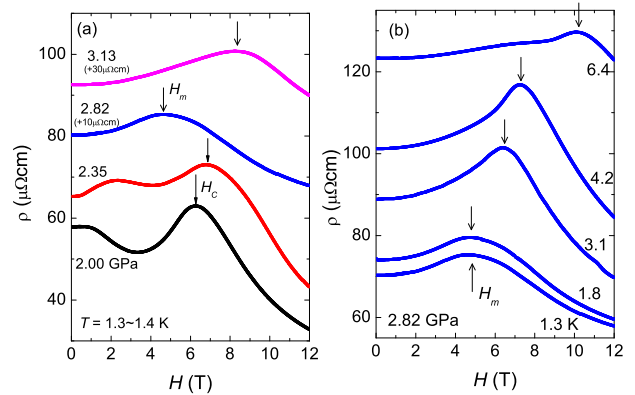}
\caption[]{(color online) Field dependences of (a) $\rho$ at $\sim1.4$ K at different pressures and (b) $\rho$ at 2.82 GPa. The critical field $H_c$, where the AFM phase is suppressed, and the MM (PM-PPM) field $H_m$ are defined by the peak of $\rho(H)$.  
}
\end{figure}

Figure 4(a) shows the field dependences of $\rho$ at different pressures measured at $\sim1.4$ K.
The broad maxima of $\rho$ indicated by arrows at 2.00 and 2.35 GPa correspond to the critical field denoted by $H_c$, at which the AFM state collapses as shown in Figs.~3(a) and 3(b).
At 2.00 and 2.35 GPa, other anomalies are observed at $1-2$ T below $H_c$.
This might be another MM transition inside the AFM phase, as observed in the doped CeRu$_2$Si$_2$.\cite{Flouquet,Aoki_CeRu2Si2,Shimizu,Haen2,Matsumoto}
In the PM region at 2.82 and 3.13 GPa, a broad peak appears at the magnetic field denoted by $H_m$, above which the anomaly at $T_0$ appears.
Figure 4(b) shows the field dependences of $\rho$ for 2.82 GPa at different temperatures.
The anomaly at $H_m$ shifts clearly to a higher field with increasing temperature. Unfortunately, the estimation of the field evolution of the $A$ coefficient was difficult owing to the presence of the hump anomaly at $T_0$.
Measurements at much lower temperatures are required for accurate estimation.

\begin{figure}[htb]
\centering
\includegraphics[width=0.9\linewidth]{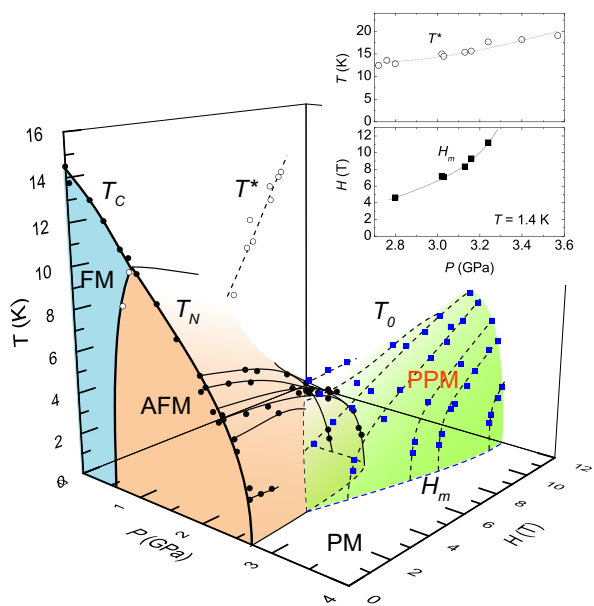}
\caption[]{(color online)  $P-T-H$ phase diagram of CeRuPO. The ordered state at ambient pressure and zero field is the FM state where the ordered moments are aligned along the $c$-axis. The AFM phase appears under pressure, and it is suppressed by applying further pressure and the magnetic field. In the high-pressure region, the MM transition into the PPM state appears, where the magnetic moments are conjectured to align along the $ab$ plane. The inset shows the pressure dependences of $T^*$ and $H_m$.
}
\end{figure}

Figure 5 shows the $P-T-H$ phase diagram for CeRuPO.
CeRuPO has neither FM QCP nor FM TCP because of the emergence of the AFM phase, but shows the MM transition into the PPM state.
The broad maximum in $\rho(H)$ at $H_m$ indicates that the MM transition is a crossover.

The MM transition near the AFM instability is reminiscent of the MM transition in doped CeRu$_2$Si$_2$.\cite{Knafo,Flouquet,Aoki_CeRu2Si2,Shimizu,Matsumoto,Aoki_Rh}
In this situation, two kinds of MM transition at $H_c$ and $H_m$ are smoothly connected with changing parameters such as pressure or doping level, but that in the case of CeRuPO seems to differ from it.
The MM transition is observed similarly in CeFePO.\cite{Kitagawa1}
It has been pointed out that the MM transition in CeFePO originates from the breakdown of the Kondo effect from the good scaling between the temperature of the susceptibility maximum, $T_{max}$, and $H_m$ in various heavy-fermion compounds.\cite{Kitagawa3}
This indicates that the energy scales of Kondo temperature and $H_m$ are comparable in CeFePO.
The $H_m \sim 4.6$ T for CeRuPO at $\sim2.8$ GPa is slightly higher than the $H_m\sim 4$ T for CeFePO, and $T^*$ is also slightly higher in CeRuPO, as shown in the inset of Fig.~1(a).
This means that the energy scales of the Kondo effect and the MM transition are comparable even in pressurized CeRuPO.
As shown in the inset of Fig.~5, however, $H_m$ increases more steeply with pressure application than $T^*$, indicating that these MM transitions cannot be explained by the Kondo breakdown scenario solely in CeRuPO.
It is conjectured that the presence of the FM correlation is an essential factor to induce the MM transition as well as for other itinerant FM systems.\cite{Valentin,Kotegawa,Aoki,Kabeya}
It will give good information on the MM mechanism whether or not $H_m$ shows good scaling with the susceptibility maximum (not $T_{max}$).\cite{Yamada_Bc}

To summarize, the electrical resistivity measurements revealed that pressurized CeRuPO shows a rich phase diagram with the switching of the magnetic ground state at zero field and the MM transition into the PPM state.
The comparison between $T^*$ and $H_m$ suggests that the energy scales of the Kondo effect and the MM transition are comparable in CeRuPO, but that the MM transition does not originates from only the breakdown of the Kondo effect.
The FM interaction is suggested to be an important factor for the MM transition, whereas the presence of the AFM instability indicates that the field evolution of the AFM interaction is also a considerable factor.
The key issue in the novel MM transition in CeRuPO is the interplay among the FM correlation, AFM correlation, and Kondo effect.

\section*{Acknowledgements}

We thank Professor Kenji Ishida for valuable discussions.
This work has been partly supported by a Grant-in-Aid for Scientific Research (No. 24340085) from the Ministry of Education, Culture, Sports, Science and Technology (MEXT) of Japan.


\begin{thebibliography}{99} 


\bibitem{Haen}
P. Haen, J. Flouquet, F. Lapierre, P. Lejay, and G. Remenyi: J. Low Temp. Phys. {\bf 67} (1987) 391.

\bibitem{Flouquet}
J. Flouquet, D. Aoki, W. Knafo, G. Knebel, T.D. Matsuda, S. Raymond, C. Proust, C. Paulsen, and P. Haen: J. Low Temp. Phys. {\bf 161} (2010) 83.

\bibitem{Knafo}
W. Knafo, D. Aoki, D. Vignolles, B. Vignolle, Y. Klein, C. Jaudet, A. Villaume, C. Proust, and J. Flouquet: Phys. Rev. B {\bf 81} (2010) 094403.


\bibitem{Aoki_CeRu2Si2}
D. Aoki, C. Paulsen, T. D. Matsuda, L. Malone, G. Knebel, P. Haen, P. Lejay, R. Settai, Y. \=Onuki, and J. Flouquet: J. Phys. Soc. Jpn. {\bf 80} (2011) 053702.

\bibitem{Shimizu}
Y. Shimizu, Y. Matsumoto, K. Aoki, N. Kimura, and H. Aoki: J. Phys. Soc. Jpn. {\bf 81} (2012) 044707.
 

\bibitem{Haen2}
P. Haen, H. Bioud, and T. Fukuhara: Physica B {\bf 259-261} (1999) 85.



\bibitem{Matsumoto}
Y. Matsumoto, M. Sugi, K. Aoki, Y. Shimizu, N. Kimura, T. Komatsubara, H. Aoki, M. Kimata, T. Terashima, and S. Uji: J. Phys. Soc. Jpn. {\bf 80} (2011) 074715. 




\bibitem{Sato}
M. Sato, Y. Koike, S. Katano, N. Metoki, H. Kadowaki, and S. Kawarazaki: J. Phys. Soc. Jpn. {\bf 73} (2004) 3418.






\bibitem{HAoki}
H. Aoki, S. Uji, A. K. Albessard, and Y. \=Onuki: Phys. Rev. Lett. {\bf 71} (1993) 2110.

\bibitem{Daou}
R. Daou, C. Bergemann, and S. R. Julian: Phys. Rev. Lett. {\bf 96} (2006) 026401.

\bibitem{Aoki_Rh}
D. Aoki, C. Paulsen, H. Kotegawa, F. Hardy, C. Meingast, P. Haen, M. Boukahil, W. Knafo, E. Ressouche, S. Raymond, and J. Flouquet: J. Phys. Soc. Jpn. {\bf 81} (2012) 034711.


\bibitem{Belitz}
D. Belitz, T. R. Kirkpatrick, and J. Rollb\"uhler: Phys. Rev. Lett. {\bf 94} (2005) 247205.


\bibitem{Yamada}
H. Yamada: Physica B {\bf 391} (2007) 42.


%\bibitem{Millis}
%A. J. Millis {\it et al.,} Phys. Rev. Lett. {\bf 88}, 217204 (2002).




%\bibitem{Yamaji}
%Y. Yamaji {\it et al.,} J. Phys. Soc. Jpn. {\bf 76}, 063702 (2007).


%\bibitem{Imada}
%M. Imada {\it et al.,} J. Phys.: Condens. Matter {\bf 22}, 164206 (2010).







\bibitem{Valentin}
V. Taufour, D. Aoki, G. Knebel, and J. Flouquet: Phys. Rev. Lett. {\bf 105} (2010) 217201.

\bibitem{Kotegawa}
H. Kotegawa, V. Taufour, D. Aoki, G. Knebel, and J. Flouquet: J. Phys. Soc. Jpn. {\bf 80} (2011) 083703.


\bibitem{Aoki}
D. Aoki, T. Combier, V. Taufour, T. D. Matsuda, G. Knebel, H. Kotegawa, and J. Flouquet: J. Phys. Soc. Jpn. {\bf 80} (2011) 094711.


\bibitem{Kabeya}
N. Kabeya, H. Maekawa, K. Deguchi, N. Kimura, H. Aoki, and N. K. Sato: J. Phys. Soc. Jpn. {\bf 81} (2012) 073706.


%\bibitem{Sidorov}
%V.A. Sidorov {\it et al.,} Phys. Rev. B {\bf 83}, 060412(R) (2011).





\bibitem{Krellner}
C. Krellner, N. S. Kini, E. M. Br\"uning, K. Koch, H. Rosner, M. Nicklas, M. Baenitz, and C. Geibel: Phys. Rev. B {\bf 76} (2007) 104418.



\bibitem{Krellner2}
C. Krellner and C. Geibel: J. Cryst. Growth. {\bf 310} (2008) 1875.




\bibitem{Kitagawa2}
S. Kitagawa, K. Ishida, T. Nakamura, M. Matoba, and Y. Kamihara: Phys. Rev. Lett. {\bf 109} (2012) 227004.


\bibitem{Kitagawa3}
S. Kitagawa, K. Ishida, T. Nakamura, M. Matoba, and Y. Kamihara: J. Phys. Soc. Jpn. {\bf 82} (2013) 033704.


\bibitem{Kitagawa1}
S. Kitagawa, H. Ikeda, Y. Nakai, T. Hattori, K. Ishida, Y. Kamihara, M. Hirano, and H. Hosono: Phys. Rev. Lett. {\bf 107} (2011) 277002.



\bibitem{Macovei}
M. E. Macovei, M. Nicklas, C. Krellner, C. Geibel, F. Steglich: Physica B {\bf 404} (2009) 2934.




\bibitem{Indenter}
T. C. Kobayashi, H. Hidaka, H. Kotegawa, K. Fujiwara, and M. I. Eremets: Rev. Sci. Instrum. {\bf 78} (2007) 023909.

\bibitem{Murata}
K. Murata, K. Yokogawa, H. Yoshino, S. Klotz, P. Munsch, A. Irizawa, M. Nishiyama, K. Iizuka, T. Nanba, T. Okada, Y. Shiraga, and S. Aoyama: Rev. Sci. Instrum. {\bf 79} (2008) 085101.



\bibitem{Lausberg}
S. Lausberg, J. Spehling, A. Steppke, A. Jesche, H. Luetkens, A. Amato, C. Baines, C. Krellner, M. Brando, C. Geibel, H.-H. Klauss, and F. Steglich: Phys. Rev. Lett. {\bf 109} (2012) 216402.






\bibitem{Yamada_Bc}
H. Yamada and T. Goto: Physica B {\bf 346-347} (2004) 109.



%\bibitem{Weickert}
%F. Weickert {\it et al.,} Phys. Rev. B {\bf 81}, 134438 (2010).















\end{thebibliography}
\end{document}